\documentclass{article}
\begin{document}

\begin{center}
\centerline{\large \ Coherent control and manifestation of inequality }
\centerline {\large \ of forward and reversed processes in optics.}
\end{center}

\vspace{3 pt}
\centerline{\sl V.A.Kuz'menko}
\vspace{5 pt}
\centerline{\small \it Troitsk Institute for Innovation and Fusion Research,}
\centerline{\small \it Troitsk, Moscow region, 142190, Russian Federation.}
\vspace{5 pt}

\begin{abstract}

The experiments with broadband down-converted light clearly show inequality 
of forward and reversed processes in optics.

\vspace{5 pt}
\end{abstract}
\vspace{12 pt}

     The concept of coherency gives good explanation of a light interference 
phenomenon. This concept is widely used now also for explanation of origin of 
majority complex nonlinear effects in optics: coherent population trapping [1],
 amplification without inversion [2], electromagnetically induced transparency 
[3], photon echo and so on. It is supposed, that interaction of coherent laser 
radiation with atoms and molecules creates a so-called coherent states, which 
have some specific properties. Some processes of constructive and destructive 
interference of such states are also supposed. As a result the discussed above 
nonlinear effects become possible.
     
     However, the physical base of such explanation is rather weak. Some 
theorists argue that the inability to measure the absolute phase of an 
electromagnetic field prohibits the representation of a laser's output as 
a quantum optical coherent state [4, 5]. In fact, it means, that using the 
concept of coherent states for explanation of nonlinear phenomena does not 
have physical sense. Such situation may be more suits for a religion, 
than for the science [6]. 
     
     From other side, we have now also the alternative explanation of origin 
of nonlinear phenomena in optics [7]. It is based on the concept of inequality 
of forward and reversed processes in optics. This concept at once assumes the 
existence of a memory of atoms and molecules about the initial state. Such 
property looks like very similar to the properties of the coherent states: 
the memory must be destroyed in collisions.
     
     The main direct and reliable experimental proof of inequality forward and 
reversed processes in optics now is the results of the study of excited 
states of polyatomic molecules in a molecular beam [8]. Polyatomic molecules 
have in absorption spectrum the so-called wide component of line, which is 
characterized by unusual combination of properties: the huge homogeneous 
spectral width of optical transition is combined with the long lifetime of 
excited states [9]. Absorption of photon is a forward process in these 
experiments. And the stimulated emission of photons is a reversed process 
(Fig.1a). The simple pump-probe experiments give the direct result that the 
spectral width of the reversed process in this case is in more than five 
orders of magnitude smaller, than the width of the forward transition. 
Accordingly, the cross-section of the reversed process is in several orders 
of magnitude greater, than for the forward transition. 
     
     The goal of this note is to discuss the recent experiments on splitting 
and mixing of photons, which also demonstrate inequality of forward and
 reversed process in optics [10, 11]. In the preceding case the memory about 
the initial state is demonstrated by molecules. In the discussed cases the 
photons demonstrate this property. In these experiments at the first step
the photons are splitted in the process of down-conversion of nanosecond 
laser pulse in two beams ($\omega_{idler}$ and $\omega_{signal}$) of 
broadband femtosecond pulses. In the second step this photons are mixed 
(Fig.1b). In one case the mixing is carried out by sum frequency generation 
in nonlinear crystal with subsequent study of spectral characteristics of 
light [10]. In other case such mixing is carried out as a two-photon 
excitation of rubidium atoms in the cell and the efficiency of mixing is 
controlled by monitoring the fluorescence intensity of excited atoms [11].
     
     Because of the spectral width of the down-converted pulses 
is very large, the mixing light also may be very wide. In this case the 
splitting of photons is a forward process. The mixing of splitted photons is 
a reversed process, but only for the case, when $\omega = \omega_1$. 
When $\omega \neq \omega_1$, the mixing of photons again will be only the 
forward process. Both experiments give the same result: the reversed process 
is much more efficient, than the forward one. The resulting radiation 
has the same narrow spectral width as the initial nanosecond laser pulse. 
This experiments quit clearly demonstrate, that the forward and reversed 
processes in optics are inequivalent.
     
     It will be also interesting to study the opposite case, when the first 
forward step is a mixing of photons and the second reversed step is a 
splitting of the mixed photons. 
     
     In conclusion, the concept of inequality of forward and reversed 
processes, probably, is a fundamental property of quantum mechanics [12]. 
This concept allows giving simple and natural explanation of physical origin 
of most effects in nonlinear optics [13].

\end{document}